\documentclass[10pt,aps,pre,twocolumn,groupedaddress,floatfix,longbibliography]{revtex4-1}

\usepackage{graphicx}
\usepackage{amsmath}
\usepackage{amssymb}
\usepackage{algcompatible}
\usepackage{hyperref}
\begin{document}

\title[Vertical drying of a suspension of sticks]{Vertical drying of a suspension of sticks: Monte Carlo simulation for continuous two-dimensional problem}

\author{Nikolai I. Lebovka}
\email[Corresponding author: ]{lebovka@gmail.com}
\affiliation{Department of Physical Chemistry of Disperse Minerals, F. D. Ovcharenko Institute of Biocolloidal Chemistry, NAS of Ukraine, Kiev, Ukraine, 03142}
\affiliation{Department of Physics, Taras Shevchenko Kiev National University, Kiev, Ukraine, 01033}
\author{Yuri Yu. Tarasevich}
\email[Corresponding author: ]{tarasevich@asu.edu.ru}
\affiliation{Laboratory of Mathematical Modeling, Astrakhan State University, Astrakhan, Russia, 414056}
\author{Nikolai V. Vygornitskii}
\affiliation{Department of Physical Chemistry of Disperse Minerals, F. D. Ovcharenko Institute of Biocolloidal Chemistry, NAS of Ukraine, Kiev, Ukraine, 03142}
\date{\today}

\begin{abstract}
The vertical drying of a two dimensional colloidal film containing zero-thickness sticks (lines) was studied by means of kinetic Monte Carlo (MC) simulations.
The continuous two-dimensional problem for both the positions and orientations was considered. The initial state before drying was produced using a model of random sequential adsorption with isotropic orientations of the sticks. During the evaporation, an upper interface falls with a linear velocity in the vertical  direction and the sticks undergo translational and rotational Brownian motions.  The MC simulations were run at different initial number concentrations (the numbers of sticks per unit area), $p_i$, and solvent evaporation rates, $u$. For completely dried films, the spatial distributions of the sticks, the order parameters and the electrical conductivities of the films in both the horizontal, $x$, and vertical, $y$,  directions were examined. Significant evaporation-driven self-assembly and stratification of the sticks in the vertical direction was observed. The extent of stratification increased with increasing values of $u$. The anisotropy of the electrical conductivity of the film can be finely regulated by changes in the values of $p_i$ and $u$. \end{abstract}

\maketitle

\section{Introduction\label{sec:intro}}

The technique of vertical drying of suspensions is frequently applied for producing colloidal crystals with many intriguing properties~\cite{Zhou2004}. However, this technique requires precise control of the drying conditions, including the concentration of particles, their size distribution, the temperature, humidity, drying time etc...~\cite{Zheng2016} Moreover, the drying process can be complicated by the formation of a skin barrier at the top of the drying film, the emergence of irregular structures and of vertical segregation as well as the development of evaporation-induced self-assembly~\cite{Routh2013,Xu2016}.

Uniformity in particle arrangement during vertical drying is controlled by solvent evaporation,  sedimentation processes,  and diffusion of the particles. It is useful to introduce the so-called P\'{e}clet number that is the ratio of the characteristic diffusive
($\tau_D$) and evaporation ($\tau_E$) times, $\mathrm{Pe}=\tau_D/\tau_E$. These times can be defined as $\tau_D=H^2/D$ and $\tau_E=H/u$, where $H$ is the initial thickness of the film in the vertical direction, $D$ is a diffusion constant for the particle and $u$ is the rate of evaporation. Uniformity in particle arrangement is expected at $\mathrm{Pe}=u H/D \ll 1$, i.e., at small $u$ or high $D$ whereas in the opposite case, for large P\'{e}clet numbers, a spatial gradient is typically observed in the density profile~\cite{Routh2013}.

Different regimes of vertical drying of suspensions filled with spherical and anisotropic colloidal particles have been extensively studied in experimental works (for a review see~\cite{Routh2013}). Experimental studies of the drying of colloidal suspensions filled with rod-like particles (with aspect ratios ranging from~4 to 15) revealed parallel orientation of the rods in several layers close to the contact line~\cite{Dugyala2015}.
Special interest has been paid to the behaviors of dried films of highly anisotropic carbon nanotubes (for a review, see~\cite{Hu2010}). These transparent and electrically conductive films present particular  interest for the production of electrodes for super-capacitors, thin film transistors, and fuel cells.

However, it is hard, experimentally, to determine many of the properties of the final dried films and the density profiles of the particles through those films~\cite{Routh2013}. Computation approaches can therefore provide valuable information, helpful for understanding the mechanisms of vertical drying and for determining the properties of the eventual dried films.

For spherical particles, Brownian, Langevin dynamics, and kinetic Monte Carlo (MC) simulations have been applied to study the vertical drying of one-, two-, three-component and polydisperse three-dimensional (3D) colloidal suspensions~\cite{Liao2000,Reyes2005,Reyes2005a,Reyes2007,Fortini2016,Fortini2017,Kameya2017}. For binary systems, either random spatial distribution or heterocoagulation were observed in dependence on differences in the surface charges of constituents~\cite{Liao2000}. The formation of ordered hexagonal/tetragonal domains and random packing at low and high evaporation rates, respectively, has also been observed~\cite{Reyes2005}. The simulation models have also included the effects of inter-particle and particle-surface interactions, and size polydispersity~\cite{Reyes2005a,Reyes2007}. At a relatively low evaporation rate, films with low porosity and low surface roughness were formed~\cite{Reyes2007}.
For polydispersed systems a dynamic stratification by size in the drying films was noted, with a layer of larger particles at the bottom and one of smaller particles on the top~\cite{Fortini2016}.  A segregation mechanism for this has been explained through the so-called colloidal diffusiophoresis effect related to the presence of a colloidal osmotic pressure gradient developed during the drying. Stratification has also been observed for ternary and polydisperse mixtures~\cite{Fortini2017}. The proposed model predicted a power-law dependence of the phoretic velocity on particle size. Kinetic MC simulations of vertical drying using 2D models at various P\'{e}clet numbers have been  performed~\cite{Kameya2017}. At $\mathrm{Pe}\gg 1$, the particles accumulated at the top of the layer whereas, at $\mathrm{Pe}\ll 1$, significant aggregation of particles in the bulk colloid could be observed and the resulting film structure became highly porous.
Vertical drying has also been studied using a hybrid simulation method based on a cell model~\cite{Gromer2015}. Various cases of hard or soft particles, different interaction potentials, and different drying conditions were tested. A linear increase of concentration gradient with increased value of $\mathrm{Pe}$, was observed although this effect was partially hindered in systems with strong repulsive interaction between the particles.

Computational studies of vertical drying have been performed much less frequently for particles with anisotropic shape.  Recently, the vertical drying of 2D colloidal films filled with anisotropic particles by means of a kinetic MC simulation has been studied~\cite{Lebovka2016}. The oversimplified lattice problem in which the particles are represented by linear $k$-mers (particles occupying $k$ adjacent sites) was addressed. The initial state of the film before drying was produced using a model of random sequential adsorption (RSA) on a square lattice, with only two restricted orientations along the lattice axes being allowed. During the drying, the upper interface falls with a linear velocity in the vertical direction, and the particles undergo translational Brownian motion. For completely dried films, the density profiles of the particles and the electrical conductivity in both directions were calculated. The simulations revealed significant evaporation-driven self-assembly together with orientation stratification of the particles with different orientations.

Note that studies of self-assembly and phase behavior in equilibrium colloidal systems filled with anisotropic particles have a fairly long history and have attracted a great deal of attention for many years. The existence of a nematic phase was firstly predicted by Onsager for a 3D continuous systems of very long rods (needles) that  interact only through the excluded volume effect~\cite{Onsager1949}.  The theory predicted a transition to this nematic phase with strong orientation ordering at some critical number density,  $\rho_n \approx 4.19$. Moreover, the coexisting isotropic and nematic phases between  $\rho_i \approx 3.29$ and  $\rho_n$ were also predicted.
Several extended approaches have subsequently been proposed and they all agree with Onsager’s estimations in the limit of low densities for sufficiently long rods~\cite{Straley1973,Lekkerkerker1984, Odijk1986, Vroege1992}.

A reduction in spatial dimensionality from 3D to 2D influences the nature of the ordered phases. The 2D version of the Onsager theory for a system of needles predicted a continuous nematic-isotropic (N--I) transition at a critical density of $\rho_i=3\pi/2\approx 4.7$~\cite{KayserJr1978}.
The 2D equivalent of the Maier--Saupe theory predicted a continuous N--I transition~\cite{Denham1980}.
A transition from disordered to a partially ordered phase was also confirmed using the MC technique for particles confined to the sites of a triangular lattice~\cite{Denham1980}.
The MC simulations demonstrated the presence of N--I transitions as particle density was increased~\cite{Frenkel1985,Bates2000}. It was shown that for the 2D systems, only a quasi long-range order with algebraic decay of the order-parameter correlation function can be realized. At relatively small densities $\rho<7$ the ordered phase became absolutely unstable with respect to disclination unbinding, but at $\rho \geq \rho_n\approx 7.25$, a 2D nematic phase with algebraic order appeared. The phase behavior of hard ellipses with an aspect ratio of $k=6$ has been studied using MC simulations~\cite{Vieillard-Baron1972}. The system exhibited two first-order phase transitions: a  solid-nematic one (at high density) and an N--I one (at a density 1.5 times smaller). These transitions were attributed to geometrical factors.
No evidence for a first order N--I transition was found using MC simulation of hard rods~\cite{Frenkel1985}. The nematic phase demonstrated algebraic order (quasi long-range order) and the occurrence of a disclination-unbinding transition of the Kosterlitz-Thouless (KT) type has been suggested.
A mean field model has predicted nematic, columnar, and crystalline order in dense systems of parallel hard rods in 2D systems~\cite{Hentschke1992}.
A density functional theory for the N--I transition in a 2D system  of rods has been developed and it predicts a continuous (second order) N--I transition~\cite{Schoot1997}.  MC simulations have been applied to study the phase behavior of continuous 2D fluid systems with spherocylinders (tapered cylinders)~\cite{Bates2000}. At high density for long rods with high aspect ratios $k\gtrsim 7$, a 2D nematic phase of the KT type occurred. Shorter rods exhibited a melting transition to an isotropically arranged phase dominated by chains of particles that aligned side-by-side.

For non-equilibrium 2D systems obtained using the RSA process, further self-assembly is possible owing to deposition-evaporation processes or to the diffusion motion of the anisotropic particles.  Several problems related to such types of self-assembly of $k$-mers have previously been discussed~\cite{Ghosh2007,Lopez2010,Loncarevic2010,Kundu2013,Kundu2013a,Matoz-Fernandez2012,Lebovka2017}. Dynamic MC simulations using a deposition-evaporation algorithm for simulation of the dynamic equilibrium of $k$-mers on square lattices have been applied~\cite{Ghosh2007}. For long $k$-mers ($k\geq 7$),  two entropy-driven transitions as a function of density, $\rho$, were revealed: first, from a low-density isotropic phase to a nematic phase with an intermediate density at $\rho_{in}$, and, second, from the nematic phase to a high-density disordered phase at $\rho_{nd}$. A lattice-gas model approach has been applied to study the phase diagram of self-assembled $k$-mers on square lattices~\cite{Lopez2010}. It has been observed that an irreversible RSA process leads to an  intermediate state with purely local orientational order while, in the equilibrium model, the nematic order can be stabilized for sufficiently long $k$-mers~\cite{Matoz-Fernandez2012,Kundu2013a}. It was also demonstrated that equilibration of the RSA structure by Brownian diffusion of the particles on a square lattice resulted in the segregation of clusters of $k$-mers with different orientations~\cite{Lebovka2017}.

Therefore, self-assembly in colloidal suspensions filled with anisotropic particles is fairly typical for both 3D and 2D systems and it can develop in different manners for continuous and lattice problems, depending on the initial state of the system and the equilibration procedure applied. This self-assembly can be even more complicated during the vertical drying processes. Nevertheless, in spite of great interest in the problem, it has  never previously been discussed for the more advanced off-lattice model taking account of the Brownian diffusion of sticks during vertical drying.

This paper analyzes the self-assembly in 2D colloidal suspensions of sticks during vertical drying. It has been assumed that the sticks are infinitely thin, and kinetic MC simulations have been applied. The initial state was produced using RSA with isotropic orientations of the sticks, after which the drying was started and the sticks were allowed to undergo both translational and the rotational diffusion. The kinetics of the changes of structure and the electrical conductivity in the horizontal ($x$) and vertical ($y$) directions have been analyzed.

The rest of the paper is constructed as follows. In Sec.~\ref{sec:methods}, the technical details of the simulations are described, all necessary quantities are defined, and some test results are given. Section~\ref{sec:results} presents our principal findings. Section~\ref{sec:conclusion} summarizes the main results.

\section{Computational model\label{sec:methods}}

The RSA model was used to produce a distribution of sticks with a desirable initial number density $\rho_i$~\cite{Evans1993}. Sticks with length of $l_s$ and zero thickness, $d_s=0$, (i.e. with infinite aspect ratio, $k=l_s/d_s=\infty$) were deposited on a plane randomly (both their positions and orientations were random) and sequentially, while their overlapping with previously placed particles was forbidden. The kinetics of RSA deposition for such systems has been studied in detail previously~\cite{Sherwood1990,Vigil1990,Ziff1990}. Note that since the sticks have zero thickness, the jamming number density $\rho_j$ is never reached for this model. However, for sticks with large but finite aspect ratios $k$, the number density is finite and increases  with $k$ as $\rho_j\sim k^{0.8}$~\cite{Vigil1989,Ziff1990}.

The length of the system was $L$ along the horizontal $x$-direction and initial height was $H_i$ along the vertical $y$-direction. In the present work, all calculations were performed using $L=H_i=32 l_s$. In our computations of vertical drying, periodic boundary conditions were applied along the $x$-axis. A zero flux boundary condition was applied at the bottom border (liquid-substrate interface). The upper border (liquid-vapor interface) moves down due to evaporation, and this interface is impenetrable for the sticks. An isotropic initial orientation of the  sticks was assumed.

The simulation of the vertical drying procedure assumed simultaneous Brownian motion of the sticks and a downward movement of the vapor-liquid interface. The Brownian diffusion of sticks was simulated using the kinetic MC procedure. At each step, an arbitrary stick was randomly chosen. The 2D translational and rotational diffusion motions were taken into consideration. For 2D translation, $D_t$, and rotational, $D_r$, the diffusion coefficients, the formulas for long, $k\gg 1$, rod-like particles were used~\cite{Li2004}
\begin{equation}\label{eq:D}
D_t= \frac{3k_BT[\ln(k)+\gamma_t]}{8\pi\eta l_s},\\
D_r= \frac{k_BT[\ln(k)+\gamma_r]}{\pi\eta l_s^3},
\end{equation}
where $k_BT$ is the thermal energy, $\eta$ is the viscosity of surrounding liquid, and $\gamma_r\approx 0.219$ and $\gamma_r\approx -0.447$ are the end correction coefficients~\cite{Broersma1960,Broersma1960a,Broersma1981}.

The ratio of the mean-square rotation $\langle\Delta \theta^2\rangle$ and mean-square displacement in the center of mass position $\langle\Delta l^2\rangle$ over the time $\Delta t$ was calculated as
\begin{equation}\label{eq:D1}
\frac{\langle\Delta \theta^2\rangle}{\langle\Delta l^2\rangle}=\frac{2D_r \Delta t}{4D_t\Delta t}\approx 4.
\end{equation}

In each MC step, displacement in an arbitrary direction by a value of $\sqrt{\langle\Delta l^2\rangle}$ or rotation by value a of $\sqrt{\langle\Delta \theta^2\rangle}$
with probabilities proportional to the corresponding diffusion coefficients $D_t$ and $D_r$, respectively, were attempted.
The value of maximum MC displacement $\sqrt{\langle\Delta l^2\rangle}$ was chosen to be small enough ($\alpha=\sqrt{\langle\Delta l^2\rangle}/l \approx 0.1-0.05$) in order to obtain satisfactory acceptance of the MC displacement~\cite{ Landau2014}. Each one time MC step, $\Delta t_{MC}$, corresponds to an attempted displacement and rotation of the total number of sticks in the system. The value of $\Delta t_{MC}$ corresponds to the real time $\tau_B \alpha^2$,  where
\begin{equation}\label{eq:Br}
\tau_B=l_s^2/(4D_t)
\end{equation}
is the Brownian time.

Time counting was started from the value of $t_{MC}=1$, being the initial moment (before drying and diffusion), and the total duration of the simulation was typically $10^6-10^7$ MC time units.

As the interface moves downward, the sticks tend to accumulate below it. Some sticks become trapped at the interface because of the effects of surface tension, and these are unable to move above the interface. Below the interface the sticks can diffuse normally.

For characterization of the processes, some  parameters were calculated during the course of the simulation:
\begin{itemize}
  \item the relative height of the film, $h=H/H_i$ ;
  \item the running number density, $\rho$;
  \item the mean order parameter of the system $S$;
  \item the distributions of local number density, $\rho(y)$, and local order parameter $S(y)$ along the vertical axis $y$;
  \item the electrical conductivities along the $x$ and $y$ axes.
\end{itemize}

The mean order parameter was calculated as
\begin{equation}\label{eq:S}
S=\frac 1 N\sum\limits_i (\cos \theta_i -1),
\end{equation}
where $\theta_i$ is the angle between the axis of the $i$-th stick and the horizontal axis $x$, and the summation runs over all $N$ particles in the system.

\begin{figure}[htbp]
  \centering
  \includegraphics[width=0.95\columnwidth]{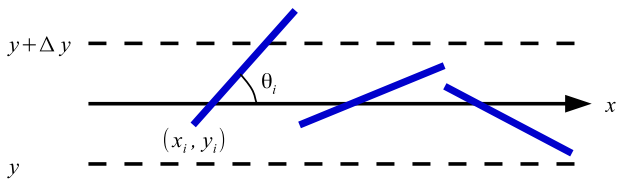}
  \caption{To the calculation of the distributions of local number density, $\rho(y)$, and local order parameter, $S(y)$, along the vertical axis $y$. Here, $y_i$ is the vertical coordinate of the center of the stick,
  $\Delta y = H/32$ is the width of the strip, $H$ is the running height of the film,
  $\theta_i$ is the angle between axis of the stick and the horizontal axis $x$.  \label{fig:Distribution}}
\end{figure}

To calculate the distributions of local number density, $\rho(y)$, and local order parameter, $S(y)$, along the vertical axis $y$, the film was divided into strips with width of $\Delta y = H/32$. In calculations of the density and order parameter profiles, those sticks with coordinates of a center of mass inside the given strip, $y \leq y_i < y +\Delta y$, were taken into account (Fig.~\ref{fig:Distribution}).

For calculation of the electrical conductivity, the  discretization scheme of the problem with a supporting square lattice of size $m\times m$ ($m=256,512,$ and $1024$) was used. The sites of supporting lattice covered by sticks were assumed as conducting and other were assumed as insulating. See Supplemental Material at [URL will be inserted by publisher] for details. The Frank--Lobb algorithm was applied to evaluate the electrical conductivity~\cite{Frank1988}. We put $\sigma_i =1$, and $\sigma_c= 10^6$ in arbitrary units for the conductivities of insulating and conducting sites, respectively. In our calculations, the two conducting buses were applied to the opposite borders of the lattice, and electrical conductivity was calculated between these buses in the horizontal, $\sigma_x$ and vertical $\sigma_y$ directions (see~\cite{Lebovka2016,Lebovka2017} for the details).

Note that our assumption of the constancy of the drying rate is a rather rough approximation. The formation of a more dense layer (crust or barrier) near the upper vapor-liquid interface can result in a restriction of the evaporation of the liquid and a decreasing value of $u$. To account for this effect, the maximum number density $\rho_m$ in the crust was calculated and a running value of the drying rate $u$ was evaluated using an exponential approximation
\begin{equation}\label{eq:uc}
u=u_i \exp(-\rho_m/\rho_c),
\end{equation}
where $u_i$ is the initial drying rate, and $\rho_c$ is a parameter (characteristic density) that controls the effect of the crust on the drying rate.

This oversimplified approximation was used only for qualitative estimations of the barrier properties of a dense layer on the drying rate. In the general case, the barrier properties can depend on details of the structure, density and thickness of the layer of sticks oriented along the evaporating interface (see~\cite{Lebovka2014} and references therein for the details). The transition from the isotropic to the ordered nematic phase was observed at $\rho\approx7$~\cite{Frenkel1985,Bates2000}. Enhancement of the barrier properties can be assumed in the nematic phase structure and, in this investigation, we put $\rho_c=7$.

\begin{figure*}[htbp]
  \centering
 \mbox{ } \hfil($a$)\hfill\qquad\qquad\qquad($b$)\hfill \mbox{ } \\
\includegraphics[width=0.45\linewidth]{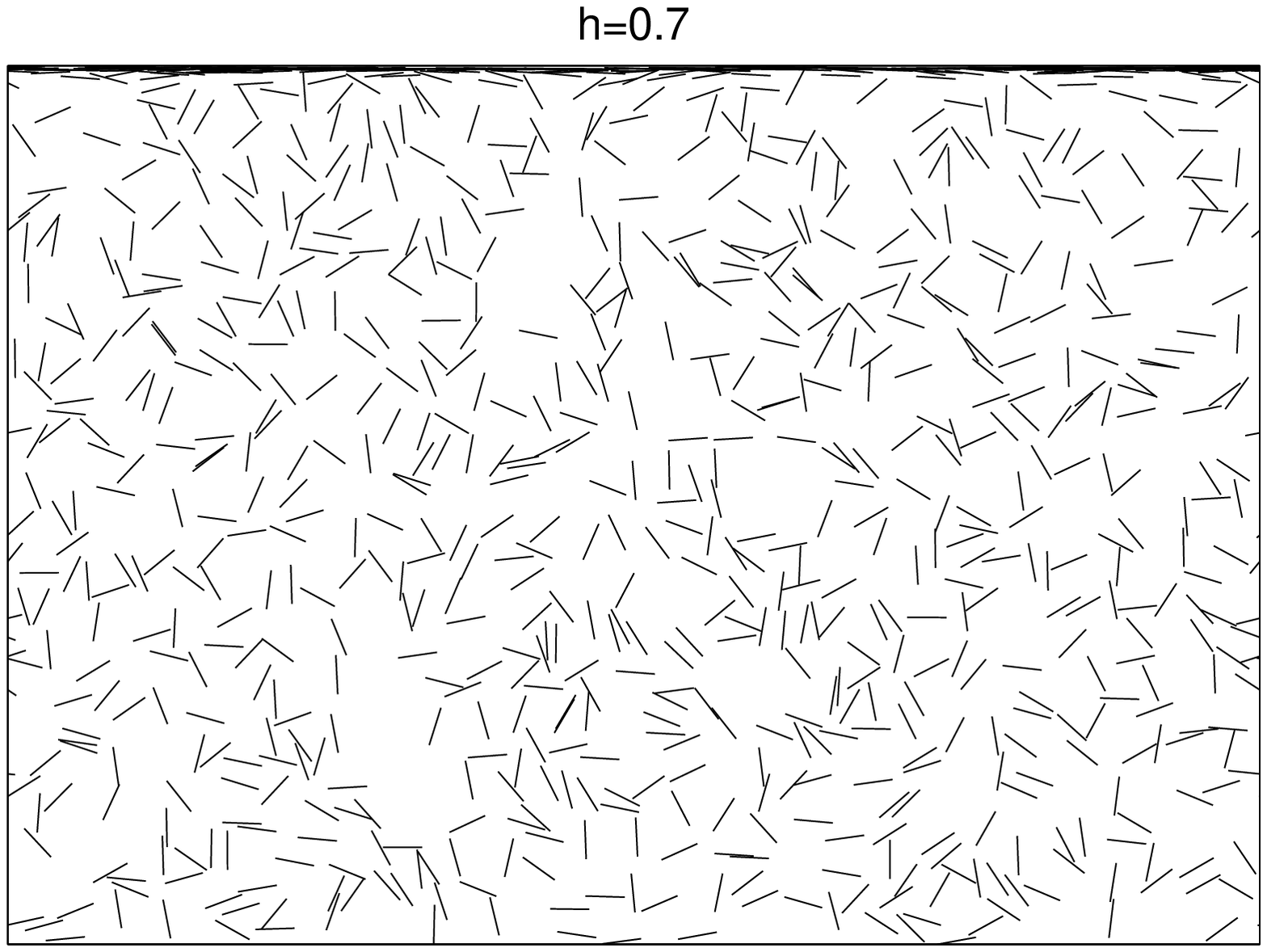}\hfill
\includegraphics[width=0.45\linewidth]{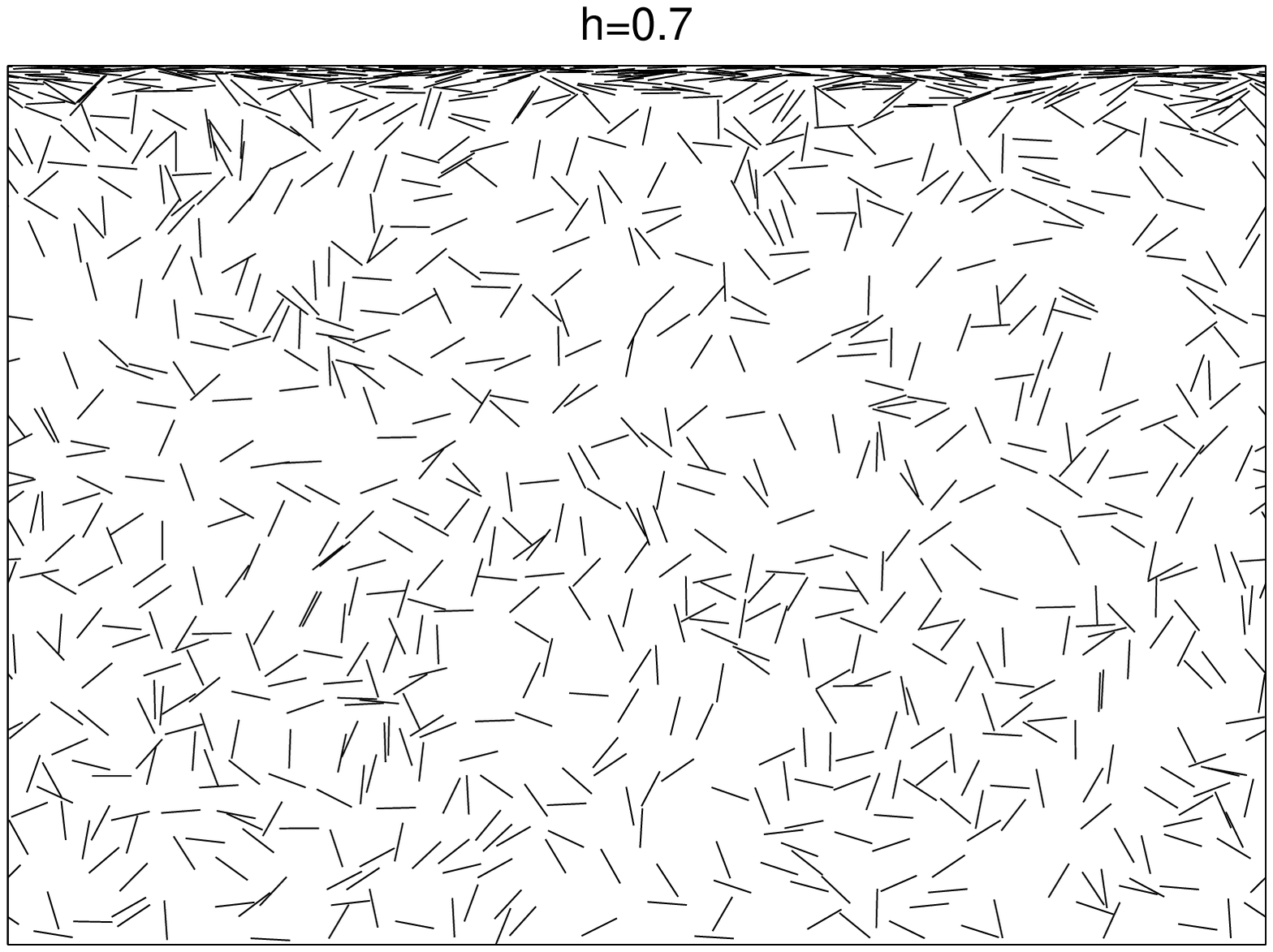}\\
\includegraphics[width=0.45\linewidth]{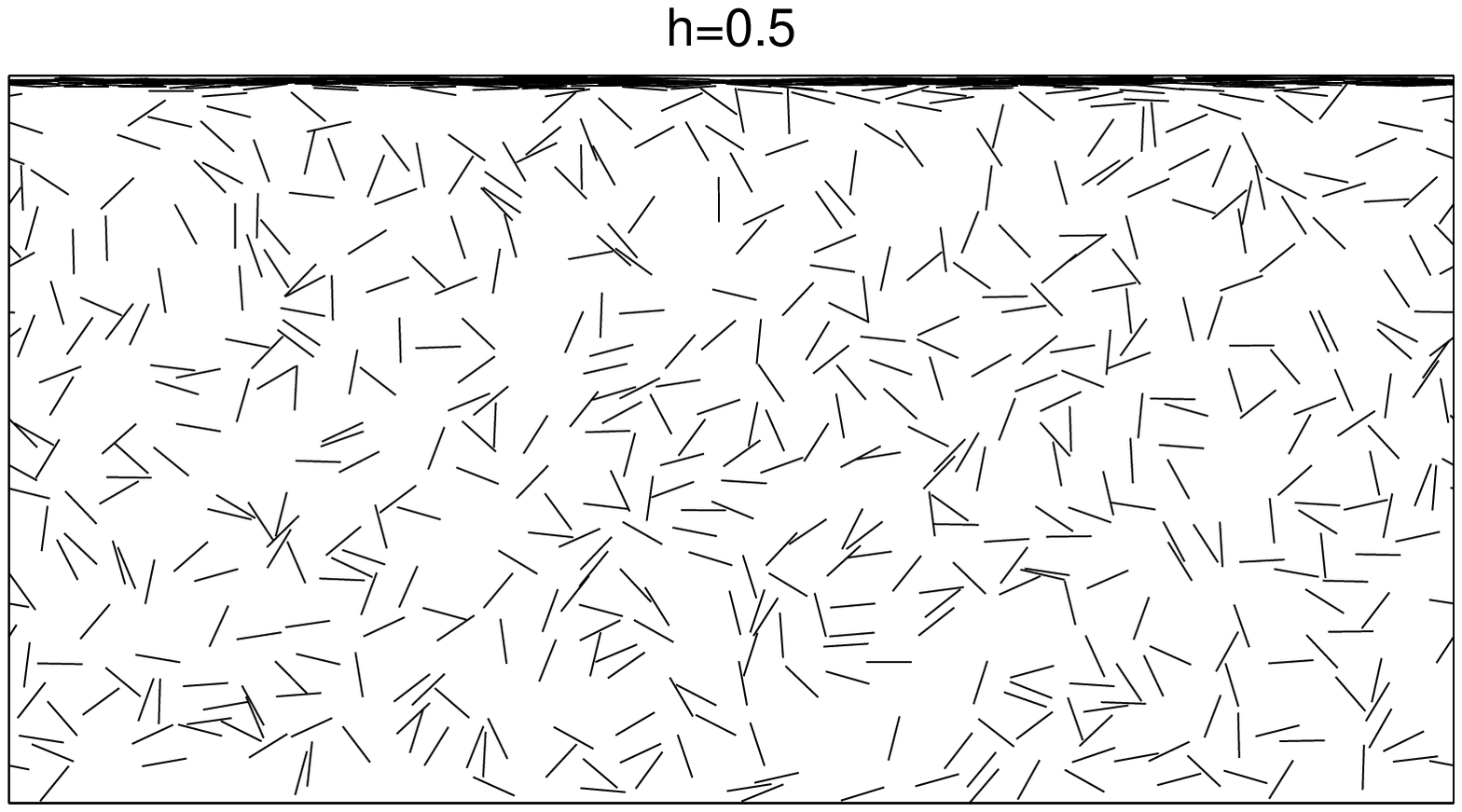}\hfill
\includegraphics[width=0.45\linewidth]{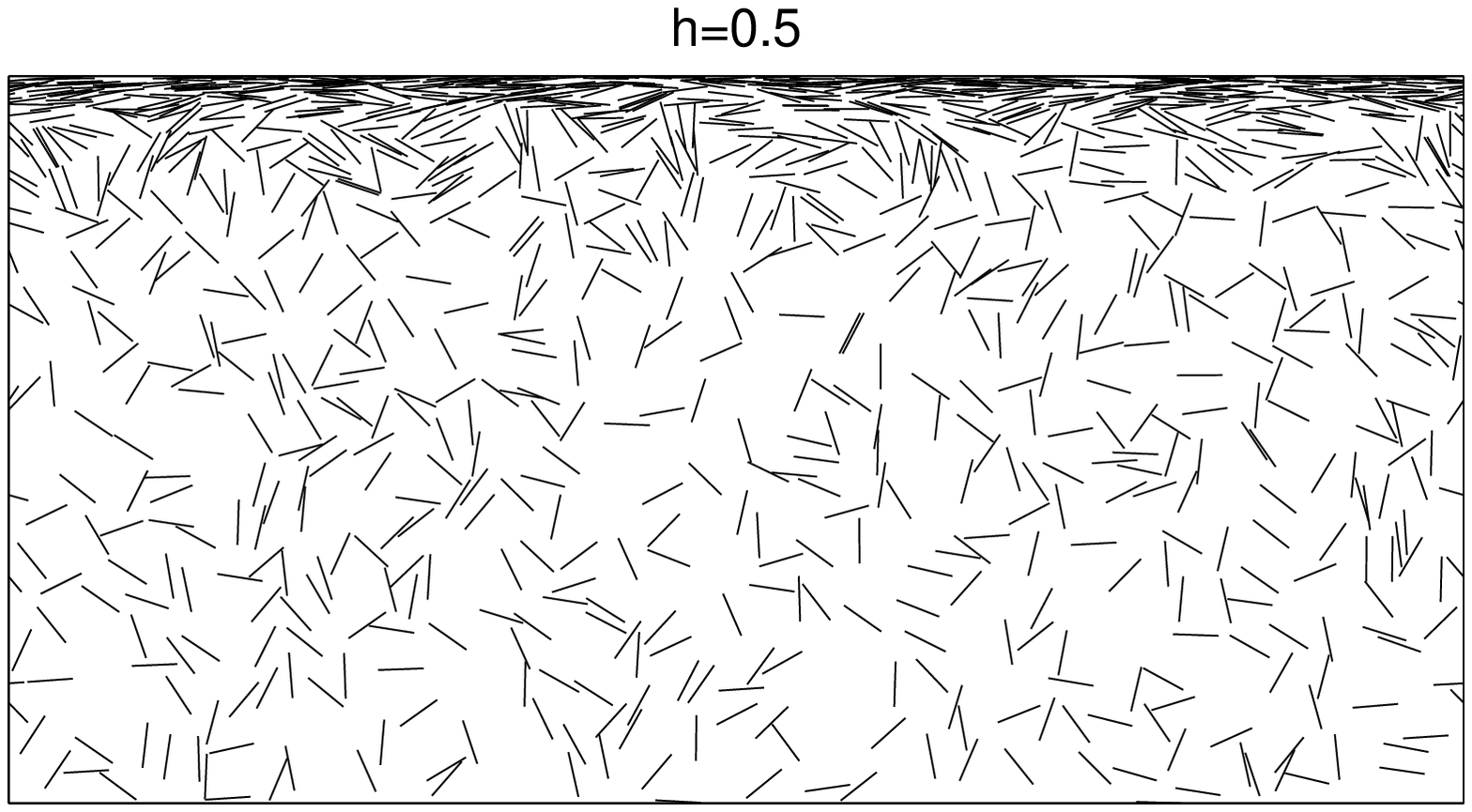}\\
\includegraphics[width=0.45\linewidth]{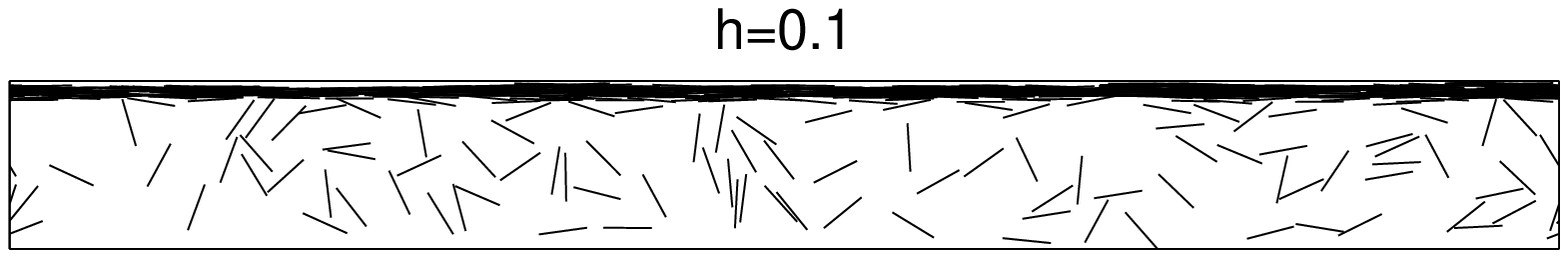}\hfill
\includegraphics[width=0.45\linewidth]{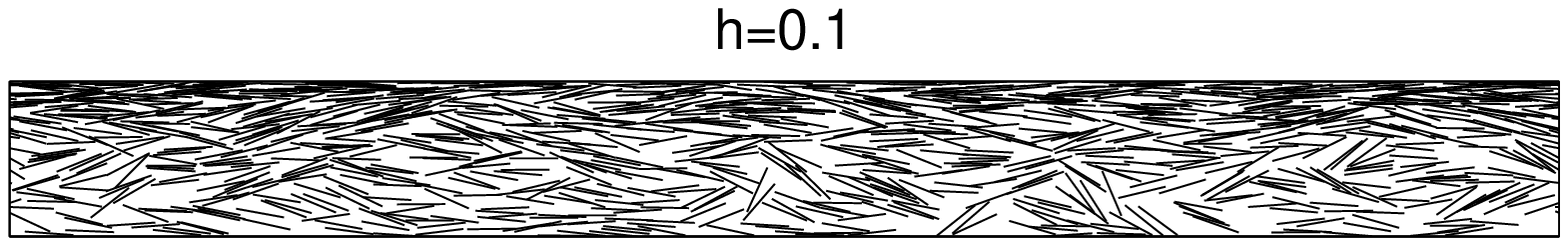}\\
  \caption{Evolution of drying patterns at different values of the relative height of the film, $h$ for a constant drying rate, $u=10^{-3}$, ($\rho_c=\infty$ in Eq. (\ref{eq:uc})) (a) and for a variable drying rate ($\rho_c=7$ in Eq. (\ref{eq:uc})) (b). The initial concentration of sticks was $\rho_i=1.0$. \label{fig:Patterns}}
\end{figure*}
\begin{figure}[htbp]
  \centering
  \includegraphics[width=1.0\columnwidth]{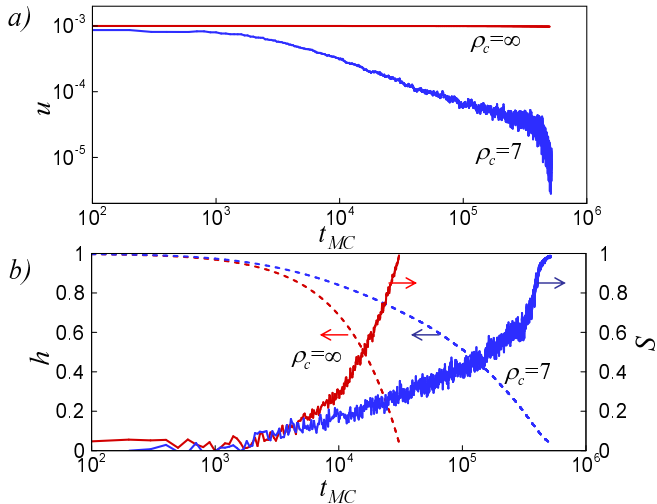}
  \caption{Drying rate, $u$ (a), relative height of the film, $h$ and average order parameter, $S$, (b) for constant and variable drying rates vs MC steps. The initial concentration of sticks was $\rho_i=1.0$. \label{fig:PatternsChar}}
\end{figure}
\begin{figure*}[htbp]
\centering
  \includegraphics[width=0.9\columnwidth]{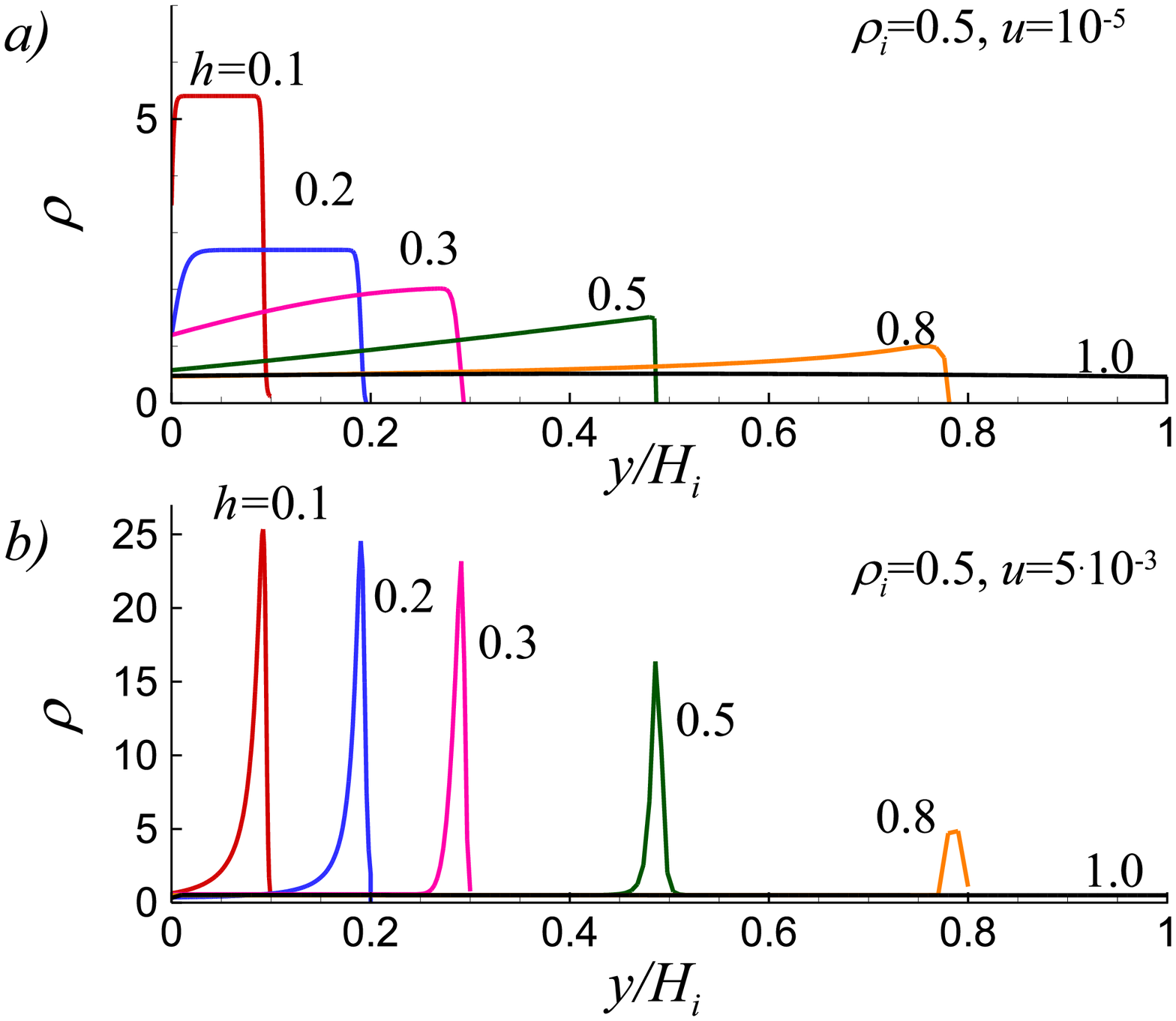}
  \includegraphics[width=0.9\columnwidth]{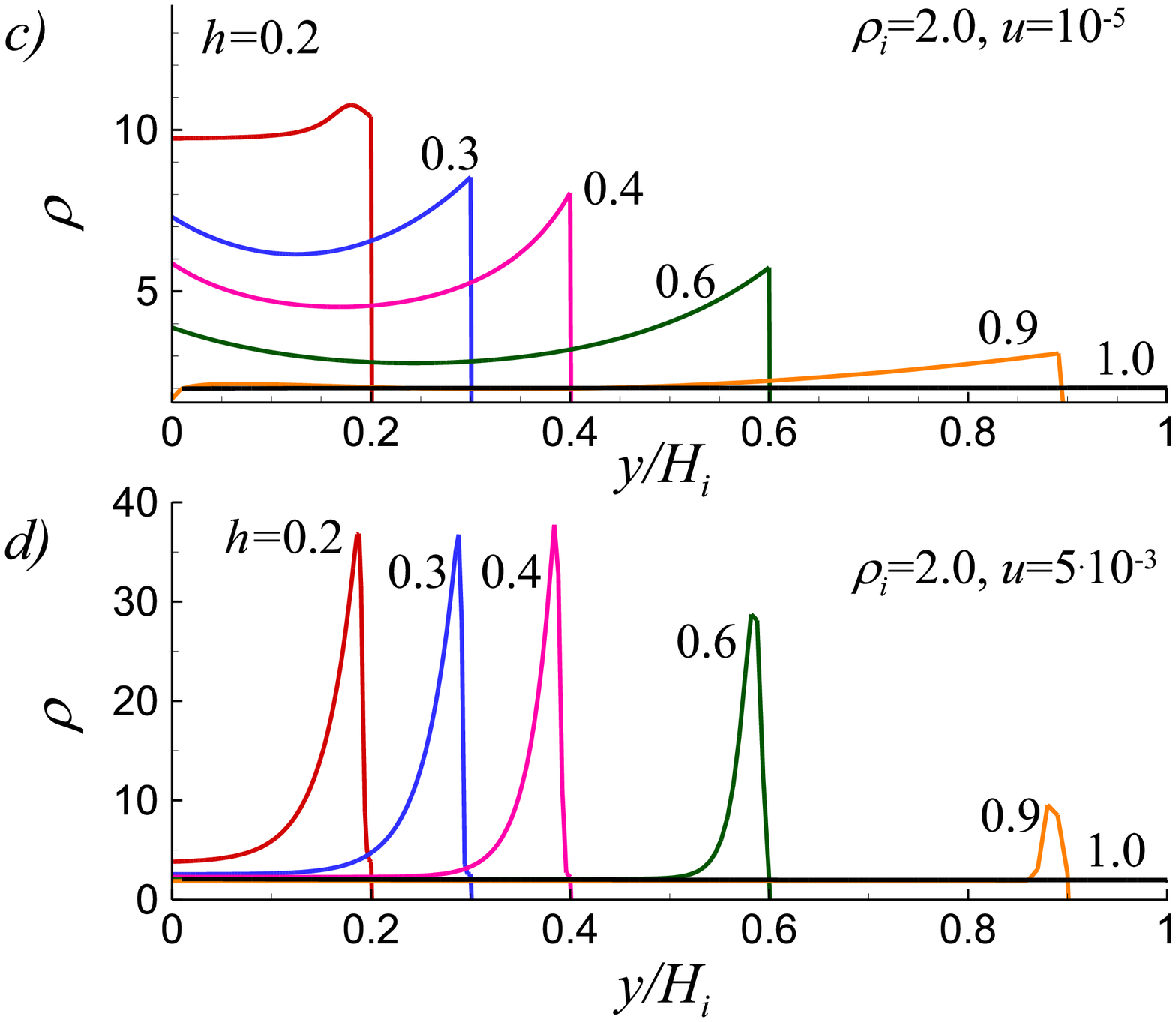}
\caption{Profiles of number density, $\rho (y/H_i)$ at different values of relative height of the film, $h$, for initial number density    $\rho_i=0.5$ and drying rate $u=10^{-5}$ (a),   $\rho_i=0.5$, $u=5\times10^{-3}$ (b),  $\rho_i=2.0$, $u=10^{-5}$ (c), and $\rho_i=2.0$, $u=5\times10^{-3}$ (d). \label{fig:f04}}
\end{figure*}

Figure~\ref{fig:Patterns} presents examples of the drying patterns at different values of the relative height of the film, $h=H/H_i$. The cases of a constant drying rate, $u=10^{-3}$, ($\rho_c=\infty$ in Eq.(\ref{eq:uc})) (a) and of a variable drying rate ($\rho_c=7$ in Eq.(\ref{eq:uc})) (b) are presented. The initial number density was $\rho_i=1.0$. Comparison of the data presented in~Fig.~\ref{fig:Patterns} allows the following preliminary conclusion to be drawn. In both cases the formation of the surface crust is evident. However, for the constant drying rate, $u=10^{-3}$ this crust is very thin and has a high number density as compared with the corresponding characteristic for the variable drying rate. The quantitative differences in the drying behavior of the systems considered above are demonstrated in Fig.~\ref{fig:PatternsChar}. For the variable drying rate, the function $u(t_{MC})$ is highly nonlinear with the most considerable drop at high values of $t_{MC}$ (Fig.~\ref{fig:PatternsChar}a). For both the constant and variable drying rates the relative height of the film, $h$, decreased while the average order parameters $S$ increased in the course of the drying (Fig.~\ref{fig:PatternsChar}b). However, the kinetics were much slower for the variable drying rate, naturally reflecting a slowdown in the drying rate. In the final stages of drying, the values of $S$ tended to one, although the nematic order was relatively homogeneous only for the variable drying rate (Fig.~\ref{fig:Patterns}a) whereas isotropic order in the bulk of the dried film was observed for the constant drying rate (Fig.~\ref{fig:Patterns}b).

The relationship between the experimental drying rate $u^r$ and the computational value, $u$, can be expressed as follows:
\begin{equation}\label{eq:Vel}
u=u^r \frac{\Delta t_{MC}}{l_s}=u^r \frac{\alpha^2 \tau_B}{l_s}.
\end{equation}

For example, using the data for water viscosity $\eta=8\times10^{4}$ Pa\,s, drying rate $u^r=3.76\times10^{-7}$~m/s at room temperature, $T=298$ K~\cite{Kroeger2007} and Eq.~\ref{eq:Br} with parameters $\alpha=1/20$ and $k=10^3$, we can obtain $u=1.34\times10^{-3}$ for $l_s=5$ $\mu$m and $u=5.37\times10^{-5}$ for $l_s=1$ $\mu$m.

The P\'{e}clet number can be evaluated as
\begin{equation}\label{eq:Peu}
\mathrm{Pe}=H^r_i u^r/D_t=\frac{4H_i}{\alpha^2}u,
\end{equation}
where $H^r_i =H_i l_s$ is the real initial height of the film.  For $H_i=32$ as used in this work, we have $\mathrm{Pe}\approx 2.56\times10^3u$.

For each given value of $\rho_i$ or $u$, the computer  experiments were repeated $100$ times, and then the data were averaged. The error bars in the figures correspond to the standard deviation of the mean. When not shown explicitly, they are of the order of the marker size.

\section{Results and Discussion\label{sec:results}}

Vertical drying processes were studied using drying rates, $u$ in the interval between $10^{-5}$ and $5\times10^{-3}$ that corresponded to the P\'{e}clet numbers in the interval between $\approx 2.56\times10^{-2}$ and $12.8$.

Figure~\ref{fig:f04} compares the profiles of number density, $\rho (y)$, for initial number density $\rho_i=0.5$ (at different drying rates, $u=10^{-5}$ ($\mathrm{Pe}\approx 2.56\times10^{-2}$) (a), and $u=5\times10^{-3}$ ($\mathrm{Pe}\approx 12.8$) (b)) and for $\rho_i=2.0$ (at $u=10^{-5}$ (c), and $u=5\times10^{-3}$ (d)).
The observed $\rho(y)$ dependencies were rather different for drying modes with small and large P\'{e}clet numbers and this was in qualitative correspondence with the previously discussed impact of the P\'{e}clet number on the spatial gradient in the density profiles~\cite{Routh2013}. For small P\'{e}clet numbers homogeneous density profiles inside the drying films were observed (Fig.~\ref{fig:f04}a,c), whereas for large P\'{e}clet numbers peaks appeared near the upper interfaces (Fig.~\ref{fig:f04}b,d) that obviously  reflected the formation of crusts.
\begin{figure}[htbp]
\centering
  \includegraphics[width=0.9\columnwidth]{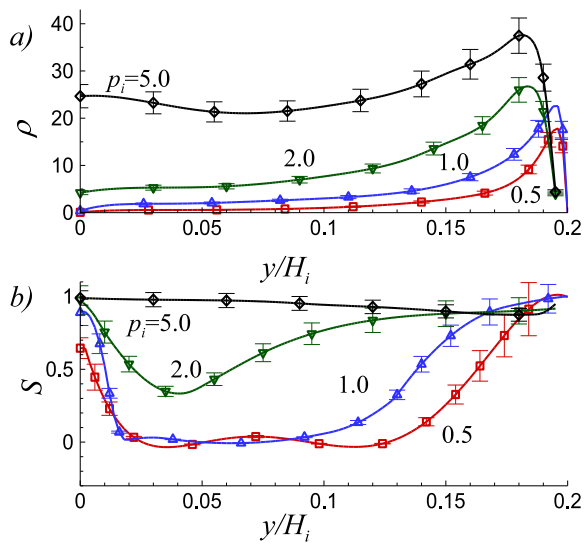}
  \caption{Profiles of number density, $\rho (y/H_i)$, (a) and local order parameter, $S (y/H_i)$, (b) at different values of initial number density, $\rho_i$, at fixed value of the relative height of the film, $h=0.2$, for the relatively high drying rate, $u=10^{-3}$.\label{fig:f05}}
\end{figure}

\begin{figure}[htbp]
\centering
  \includegraphics[width=0.9\columnwidth]{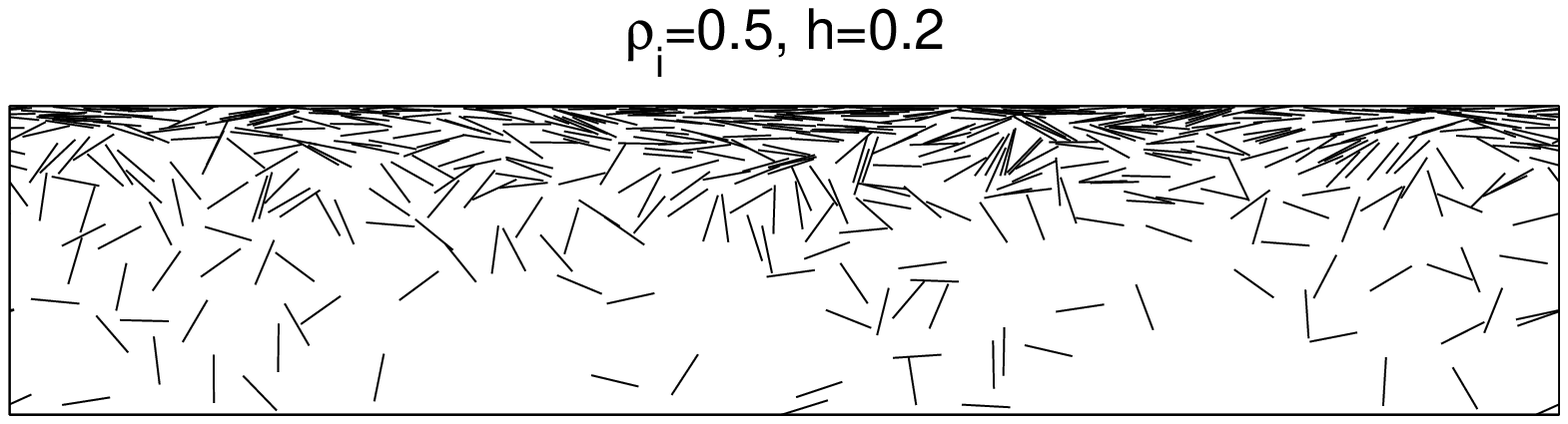}\\
  \includegraphics[width=0.9\columnwidth]{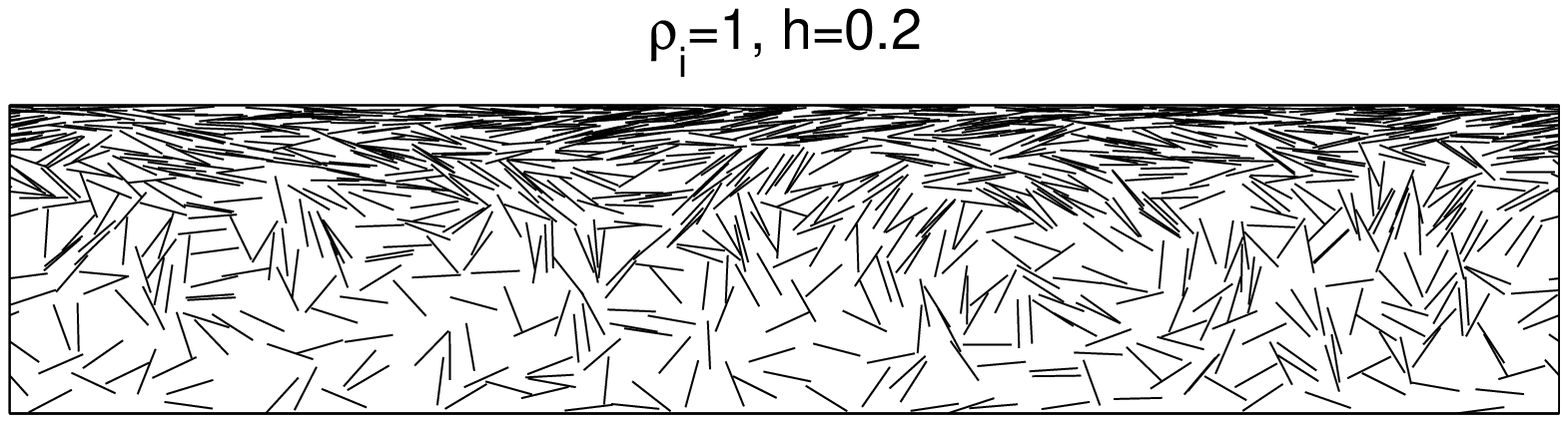}\\
  \includegraphics[width=0.9\columnwidth]{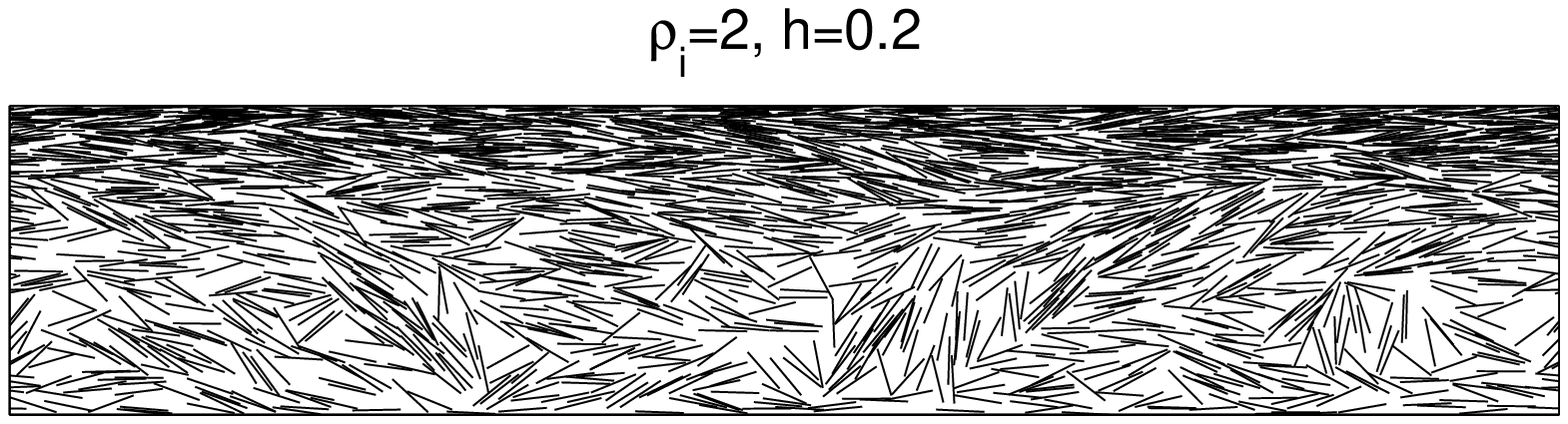}\\
  \includegraphics[width=0.9\columnwidth]{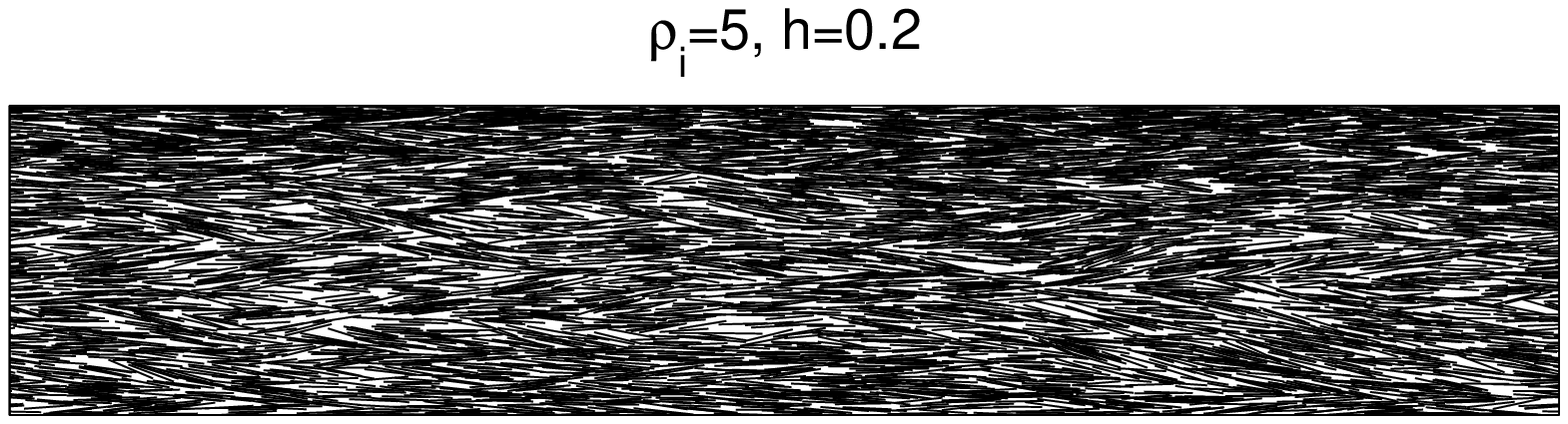}\\
  \caption{Comparison of the drying patterns at different values of the initial number density, $\rho_i$, at a fixed value of the relative height of the film, $h=0.2$, for the relatively high drying rate, $u=10^{-3}$. See Supplemental Material at [URL will be inserted by publisher] for animations of (b) and (d).\label{fig:f06abpatterns}}
\end{figure}

This conclusion is supported by the data on the profiles of the number density, $\rho (y)$, and local order parameter, $S(y)$ (Fig.~\ref{fig:f05}a,b). Here, the profiles are compared  at different initial number densities, $\rho_i$,  at the fixed value of the relative height of the dried film, $h=0.2$, for a relatively high drying rate, $u=10^{-3}$. The average number density at the given value of $h$ can be evaluated as $\rho=\rho_i/h$. The corresponding patterns for different initial number densities, $\rho_i$, are presented in (Fig.~\ref{fig:f06abpatterns}). We see that at the highest value of $\rho_i = 5$ insignificant densifications were observed near both the upper and bottom interfaces while almost  ideal ordering with $S\approx 1$ was observed inside the dried film. At smaller initial densities, the densification (crust) and significant nematic ordering were only observed near the upper interface. Less significant orientational ordering was also observed near the bottom interface (Fig.~\ref{fig:f05}b) and it decreased with decreasing values of $\rho_i$.

\begin{figure*}[ht]
  \centering
  \includegraphics[width=0.9\columnwidth]{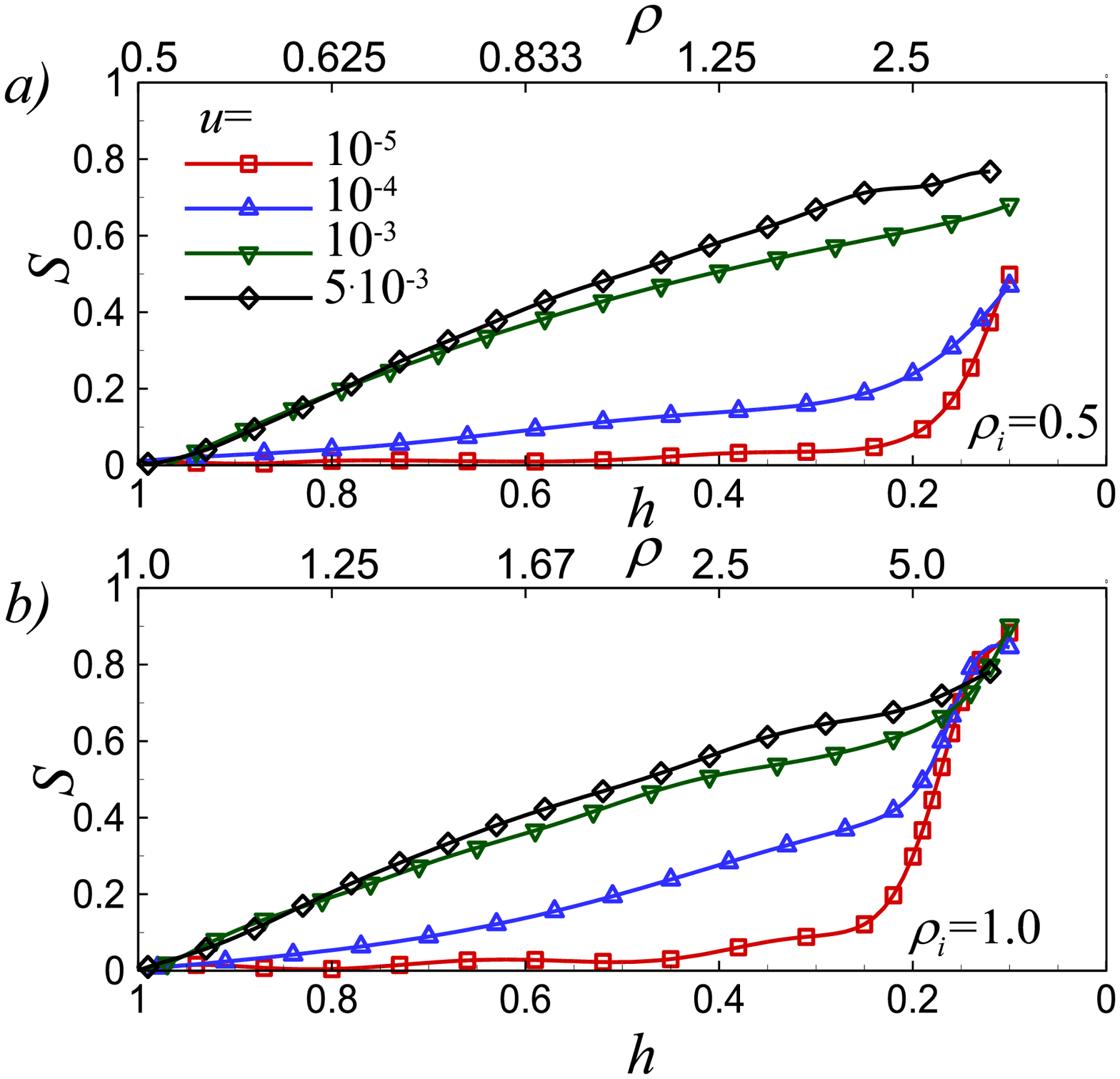}
  \includegraphics[width=0.9\columnwidth]{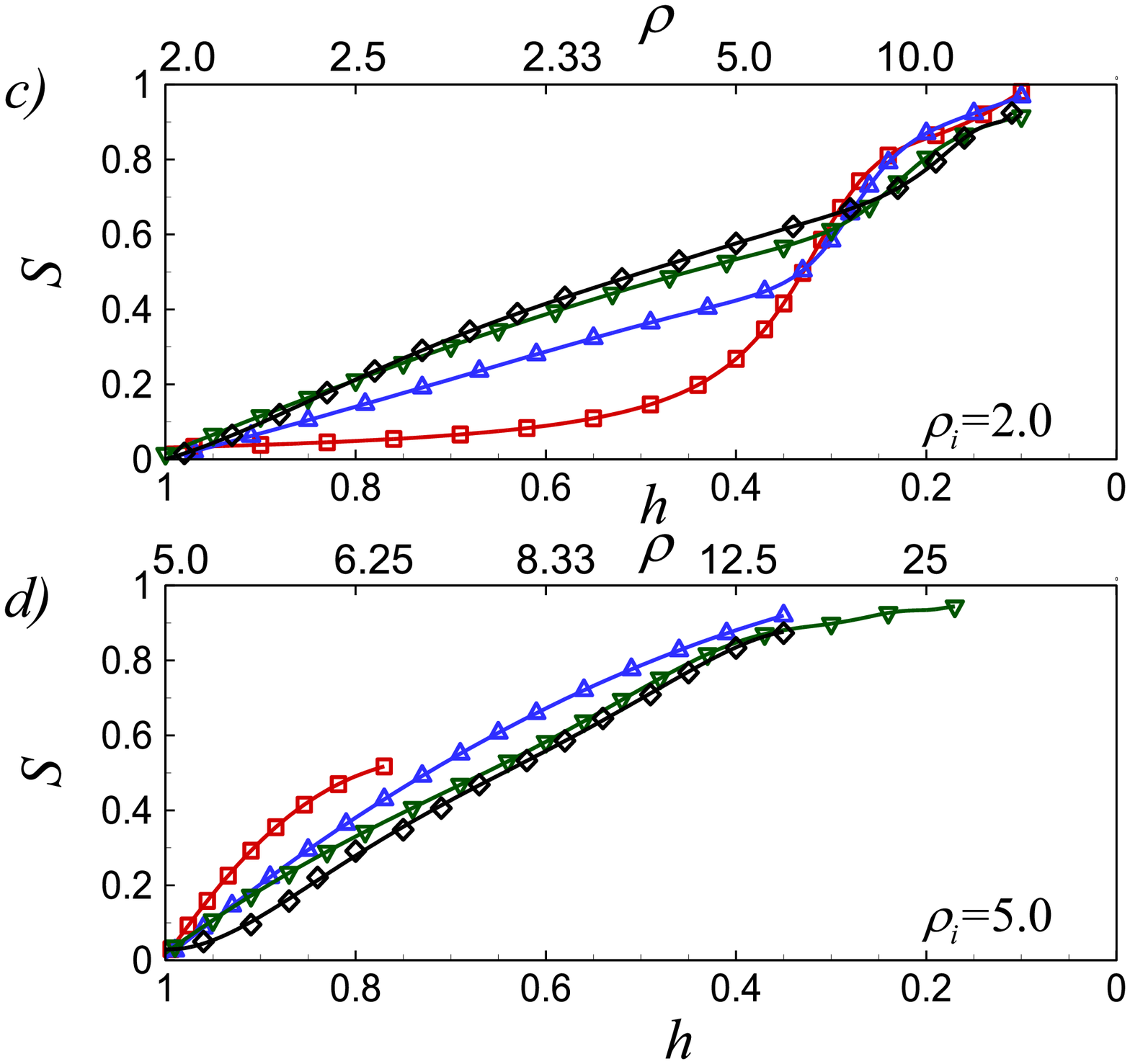}
  \caption{Mean order parameter, $S$, versus the relative height of the film, $h$, at different drying rates, $u$, and initial number densities, $\rho_i=0.5$ (a), $1.0$ (b), $2.0$ (c), and $5.0$ (d). The corresponding densities, $\rho$, are also presented on the upper horizontal axes.\label{fig:f03}}
\end{figure*}

Figure~\ref{fig:f03} presents examples of the mean order parameter, $S$, versus the relative height of the film, $h$, at different drying rates, $u$, for initial number densities, $\rho_i=0.5$ (a), $1.0$ (b), $2.0$ (c), and $5.0$ (d). Development of the drying process corresponds to the decrease in value of $h$ and the increase in  value of $\rho$.

The order parameter, $S$, always increased with increase of $h$. The observed $S(h)$ dependencies were also rather different for drying modes with small and large P\'{e}clet numbers. At relatively small initial concentrations of sticks ($\rho_i =0.5$, $\rho_i =1.0$, and $\rho_i = 2.0$, Fig.~\ref{fig:f03}a,b,c) an increase of $u$ resulted in a more pronounced  increase in the value of $S$ with decreasing value of $h$.

For high evaporation rates, $u=5\times10^{-3}$, and $u=10^{-3}$, a continuous increase in the value of $S$ with decreasing value of $h$ was observed and this increase evidently reflected the formation of the crust layer near the vapor-liquid interface. For the small evaporation rates, $u=10^{-4}$, and $u=10^{-5}$, a noticeable  increase in the value of $S$ was only observed in the final stages of drying. 
This could reflect the formation of dried layers with bulk ordering of the sticks. 
However, at the highest initial concentration of sticks, ($\rho_i = 5$, Fig.~\ref{fig:f04}d) the opposite behavior was observed and an increase of $u$ resulted in a less pronounced  increase in $S$ with decreasing value of $h$. Such behavior could reflect the effects of bulk ordering at high concentrations  of sticks.

\begin{figure*}[ht]
\centering
\includegraphics[width=0.9\columnwidth]{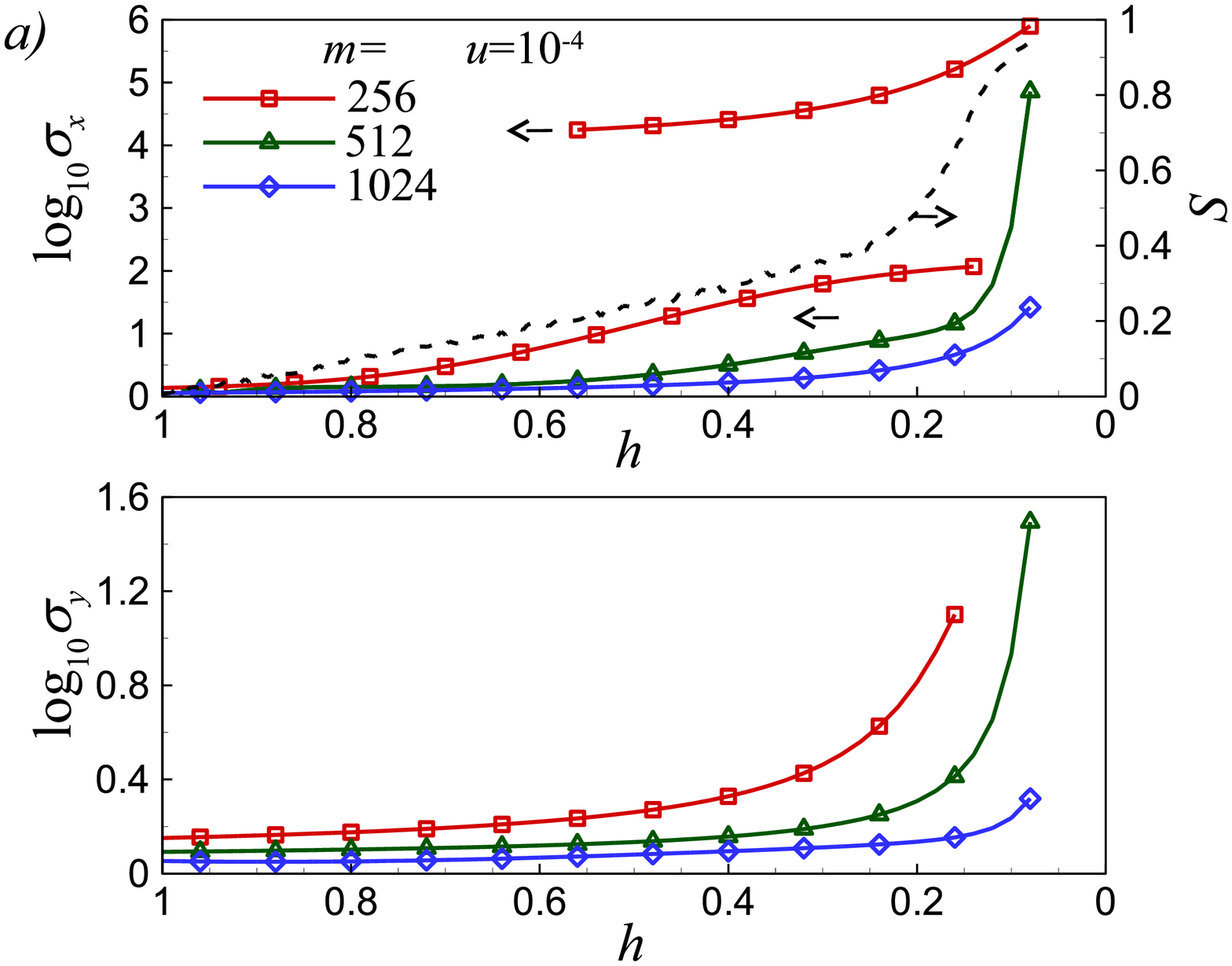}
\includegraphics[width=0.9\columnwidth]{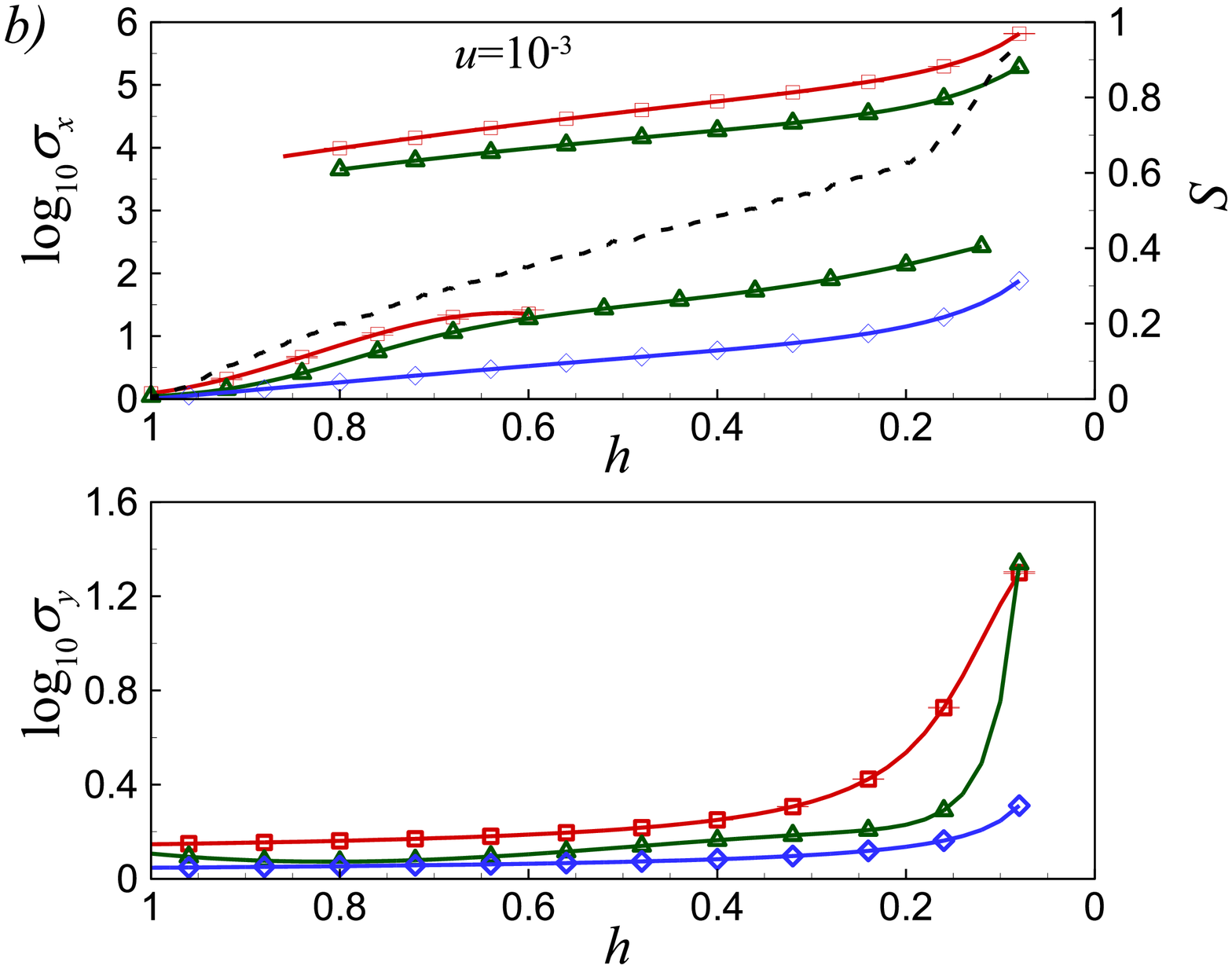}
 \caption{Electrical conductivity in the horizontal, $\sigma_x$, and vertical, $\sigma_y$, directions versus the relative height of the film, $h$, at an initial density of $\rho_i=0.1$ and different drying rates, $u=10^{-4}$, (a) and $u=10^{-3}$, (b). The data are presented for different sizes of the supporting square lattice, $m=256,512,$ and $1024$. The dashed lines correspond to the evolution of order parameter, $S$, during the drying.\label{fig:conduct}}
\end{figure*}

Figure~\ref{fig:conduct} presents examples of the electrical conductivity in the horizontal, $\sigma_x$, (a,c) and vertical, $\sigma_y$, (b,d) directions versus the relative height of the film, $h$. The initial density was $\rho_i=0.1$ and different drying rates, $u=10^{-4}$, (a,b) and $u=10^{-3}$, (c,d) were considered.
The data are presented for different sizes of the supporting square lattice, $m=256,512,$ and $1024$.
In fact, the use of the supporting square lattice for the calculation of electrical conductivity is equivalent to the rasterization of the structure of the infinitely thin sticks and substitution of them by the sticks with a finite aspect ratio of the order of $k\approx m/32$.
The system with infinitely thin sticks corresponds to that with very large values of $m/32\gg 1$.

The values of $\sigma_x$ and $\sigma_y$ increased with decreasing value of $h$ in the course of drying as this naturally reflects the changes in stick density in the film. Moreover, the calculated data pointed to the presence of significant anisotropy of electrical conductivity in the drying films with $\sigma_x>\sigma_y$. Such anisotropy clearly reflected the development of the preferential orientation of the sticks along the horizontal direction, $x$, during the drying (see, the dashed lines in Fig.~\ref{fig:conduct}a,b for the dependency of the order parameter, $S(h)$). In the vertical direction, $y$, all the dependencies $\sigma_y$ for different values of $m$ and $u$ were continuous.

However, in the direction $x$ the dependencies $\sigma_x$ can be discontinuous  in some intervals of the $h$ values (e.g., for $u=10^{-4}$, $m=256$, Fig.~\ref{fig:conduct}a, and for $u=10^{-3}$, $m=256$, or $m=512$, Fig.~\ref{fig:conduct}b). The shaded area in Fig.~\ref{fig:conduct}a represents an example of such a discontinuity, with two branches for the electrical conductivities. The upper and lower branches correspond to the formation and decomposition of percolation networks between the sticks with a small aspect ratio $k = 8$, while with $k = 16, 32$, the discontinuities were not observed. The effects of such a discontinuity were even more pronounced for a higher drying rate (Fig.~\ref{fig:conduct}a) as compared with the smaller one in Fig.~\ref{fig:conduct}b. Such behavior may  reflect the impact of the drying rate on the ordering of the sticks and the formation of the crust layer near the upper vapor-liquid interface at a high evaporation rate, $u=10^{-3}$ (i.e., at $\mathrm{Pe}\gg1$).

\section{Conclusion\label{sec:conclusion}}
The continuous 2D model of the vertical drying of a suspension of infinitively thin sticks was studied by MC simulation. The initial homogeneous and isotropic state before drying was produced using the RSA model. As a result of  evaporation, the upper interface moved down and the sticks underwent  both  translational and rotational Brownian motions. The restriction of evaporation due to the densification of the system near the upper interface was taken into account.  The data revealed the presence of evaporation-driven and diffusion-driven self-assemblies and stratification of the sticks in the vertical direction. The sticks tend to orient along the horizontal direction in several layers close to both the upper (liquid-vapor) and bottom (liquid-solid wall) contact lines.  The effects were dependent on the P\'{e}clet number and the initial number density. For small P\'{e}clet numbers, homogeneous density profiles inside the drying films were observed, whereas for large P\'{e}clet numbers,  peaks appeared near the upper interfaces that evidently reflected the formation of crusts. The results obtained were in qualitative correspondence with the previously discussed impact of P\'{e}clet number on the spatial gradient in density profiles~\cite{Routh2013}.  At small initial density ($\rho_i =0.5$), the densifications (crusts) were only observed near the upper boundary, but significant nematic orderings were still observed near both the upper and bottom boundaries. However, for dense systems ($\rho_i = 5$), insignificant densifications were observed near both the upper and bottom interfaces and practically ideal ordering with $S\approx 1$ was observed inside the film. Our data has demonstrated the presence of significant anisotropy of electrical conductivity in the drying films with $\sigma_x>\sigma_y$. Such anisotropy evidently reflects the development of a preferential orientation of sticks in the horizontal direction, $x$, during the drying. This anisotropy can be finely regulated by changing the P\'{e}clet number and initial number density. These results present an interesting possibility for prediction of the electrical conductivity of dried films, obtained by the drying of suspensions filled with highly anisotropic nanoparticles.

\section*{Acknowledgements}
We acknowledge the funding from the National Academy of Sciences of Ukraine, Project No.~43/17-H (N.I.L., N.V.V.) and the Ministry of Education and Science of the Russian Federation, Project No.~3.959.2017/4.6 (Yu.Yu.T.).

\bibliography{Drying}

\end{document}